\documentclass[a4paper,11pt]{article}

\usepackage[utf8]{inputenc}
\usepackage{amsmath, amssymb}
\usepackage{graphicx}
\usepackage{hyperref}
\usepackage{caption}
\usepackage{subcaption}
\usepackage{bm}
\usepackage{hhline, multirow, booktabs}
\usepackage{comment}

\usepackage[sorting=none,style=numeric,backend=biber]{biblatex}

\newcommand{\vect}[1]{\boldsymbol{#1}}

\title{Forecasting chaotic dynamic using hybrid system}
\author{Michele Baia${}^1$, Tommaso Matteuzzi${}^{2,3}$, Franco Bagnoli${}^{2,3}$\\
\begin{minipage}{0.9\textwidth}
~\\[.2cm]
1) Dept.of Department of Architecture and CSDC, University of Florence, via della Mattonaia 8, 50121 Firenze, Italy\\[.2cm]
2) Dept.of Physics and Astronomy and CSDC, University of Florence, via G. Sansone 1, 50019 Sesto Fiorentino,  Italy\\[.2cm]
3) INFN, Sect. Florence
\end{minipage}}
\date{\today}

\begin{document}

\maketitle

\begin{abstract}

The literature is rich with studies, analyses, and examples on parameter estimation for describing the evolution of chaotic dynamical systems based on measurements, even when only partial information is available through observations.

However, parameter estimation alone does not resolve prediction challenges, particularly when only a subset of variables is known or when parameters are estimated with significant uncertainty. In this paper, we introduce a hybrid system specifically designed to address this issue. The method involves training an artificial intelligent system to predict the dynamics of a measured system by combining a neural network with a simulated system. By training the neural network, it becomes possible to refine the model’s predictions so that the simulated dynamics synchronize with the actual system dynamics.

After a brief contextualization of the problem, we introduce the hybrid approach employed, describing the learning technique and testing the results on two chaotic systems inspired by atmospheric dynamics in measurement contexts. Although these systems are low-dimensional, they encompass all the fundamental characteristics and predictability challenges that can be observed in more complex real-world systems.

\end{abstract}

\section{Introduction}

The study of chaotic systems spans across multiple disciplines, incorporating principles of dynamical systems, nonlinear dynamics, and complex systems. Chaos theory is central to the explanation of non-periodic and unpredictable behaviors across various natural and engineered systems. A key feature of chaotic dynamics is its extreme sensitivity to initial conditions, where small discrepancies between two initial conditions can exponentially amplify over time, making long-term predictions unreliable while preserving short-term predictability. This dual nature of chaos poses significant challenges but also provides valuable insights into the behavior of complex systems.

Time series forecasting, which involves predicting future system states from historical data, is a critical tool for understanding such dynamics. The strength of traditional mathematical models lies in their direct interpretability in physical terms, however, they often struggle with uncertainties in system parameters and measurement noise, especially in chaotic regimes. In contrast, data-driven methods, particularly those employing neural networks, have shown remarkable progress in capturing intricate patterns and nonlinear relationships within time series data \cite{HEWAMALAGE2021388, Sakib2024, li2019water}.

Deep learning architectures \cite{goodfellow2016deep}, such as Convolutional Neural Networks (CNNs) and Recurrent Neural Networks (RNNs), have proven effective for tasks ranging from image recognition to sequential data analysis. Recent advances in neural architectures, such as the transformer network \cite{vaswani2017attention}, have further improved long-range dependency modeling, but limitations like the loss of temporal scale information remain \cite{zeng2023transformers}. Studies have addressed these gaps by proposing simplified models, such as LSTF-Linear \cite{wang2024tcn}, optimized for time series forecasting tasks.

In chaotic time series forecasting, neural networks like Long Short-Term Memory (LSTM)\cite{hochreiter1997LSTM} units and Gated Recurrent Units (GRUs)\cite{Cho2018GRU} have demonstrated superior performance. Research shows these models can capture the complex attractor structures inherent in chaotic systems \cite{cestnik2019inferring, cannas2002neural}, enabling more extended prediction horizons compared to traditional methods \cite{farmer1987predicting, pan2008identification, barbosa2022learning}. Furthermore, hybrid approaches that combine neural networks with modeling techniques have emerged as a promising paradigm \cite{pathak2018hybrid, kashinath2021physics}. These models exploit the structured insights of physics-based equations while leveraging neural networks to learn corrections, enhancing robustness and accuracy.

Recent studies have demonstrated the effectiveness of hybrid approaches in various contexts. For instance, Physics-Informed Neural Networks (PINNs) \cite{raissi2017physics} integrate the structure of partial differential equations into the learning process, enhancing the accuracy of predictions even in chaotic systems \cite{cai2021physics}. Neural Ordinary Differential Equations (NODEs) \cite{chen2018neural} extend this concept by embedding neural networks within the integration process, enabling adaptive modeling of nonlinear dynamics. Additionally, neural network-trained solution bundles enable efficient trajectory prediction and uncertainty quantification in chaotic systems, facilitating tasks such as Bayesian parameter inference \cite{seleznev2019bayesian, radev2020bayesflow}.

These approaches have been applied successfully to benchmark chaotic systems such as the Lorenz '63, Mackey-Glass, and Rössler systems, demonstrating improved long-term prediction capabilities compared to traditional methods. They also hold potential for real-world applications, including astrophysics, weather forecasting, and complex biological systems, where both short-term accuracy and long-term stability are critical \cite{zhang2003time, aslam2023multi, zhu2023time, waqas2024critical}. For instance, hybrid models combining RNNs with techniques like empirical mode decomposition (EMD) have outperformed standalone methods in financial forecasting \cite{cao2019financial}. Similarly, hybrid architectures such as GRU-LSTM combinations and deep temporal modules have excelled in predicting wind power \cite{hossain2020hybrid} and traffic flow, demonstrating their adaptability across domains \cite{gao2020short}.
By addressing the limitations of purely physics-based or purely data-driven methods, these hybrid models provide a robust framework for trajectory prediction and uncertainty propagation, offering new avenues for the study and control of chaotic dynamics.

On the other hand, problems such as the statistical representation of extreme events, the difficulty modeling long-term temporal dependencies, and the lack of flexibility towards data external to the domain on which they were trained \cite{shumailov2024ai} (e.g. covariate shift and concept drift),  make neural networks still unreliable for operational tasks, such as prediction in the atmospheric and meteorological fields.

A further advantage of the mathematical description of dynamic evolution, however, is that, when the function $\vect{f}(\vect{x}, \vect{q})$, where $\vect{x}$ and $\vect{q}$ represent, respectively, the state vector and the vector of the model parameters, correctly represents the dynamics of the problem, through the analysis of the dynamical equations it is possible to estimate the correct statistics, return times of rare events and prediction of unobserved dynamics.
But, even assuming that we know the dynamic system
\begin{equation}
 \dot{\vect{x}} = \vect{f}(\vect{x}, \vect{q}),
 \label{eq:DS}
\end{equation}
where the dot operator denotes the time derivative, the problem of parameter estimation remains.

Also, in general, the parameters can be estimated with a finite precision due to measurement noise and inherent limitations in parameter estimation. \\
In chaotic systems, even small discrepancies between the true and modeled parameters lead to an exponential divergence between the observed trajectory and the trajectory predicted by \( \vect{f}\).

For these reasons, a further line of research regarding the use of intelligent systems combined with concepts of synchronization and data assimilation, has moved towards the  use of neural network and the assimilation of measurements to correct prediction output iteratively and  match the forecasting to the observed data \cite{bonavita2020machine, wikner2021using, penny2022integrating, howard2024machine,  cheng2024efficient}.

This paper explores a similar hybrid approach, combining recurrent neural networks with differential equations, described by the function \( \vect{f} \), to predict chaotic trajectories, correct for model errors, and interpolate states where data are unavailable.
The neural network module compensates for this divergence by learning to correct systematic errors, thereby enabling accurate trajectory predictions over longer time horizons.

The method proposed combines a neural network with a simulated system to predict the dynamics of the measured system. The neural network is trained to correct the simulated model, thereby improving alignment with the real system. This approach is particularly useful in scenarios where partial measurements or uncertain parameters do not allow accurate predictions.

The paper is organized as follows. In the next section we introduce the problem, setting it in a short-time system prediction perspective starting from a set of measurements. We will then introduce the hybrid system, the proposed NN architecture and how it can be trained to force the dynamics estimated by the model $\vect{f}(\cdot, \vect{q}_{M})$ to follow the true one.
The third section provides a brief overview of the benchmark models used to evaluate our hybrid system, followed by the presentation of training results. We demonstrate how this system can achieve long-term predictions even when only partial measurements of the system are available.

A concluding section follows with comments and future perspectives.

\begin{figure}
    \centering
    \includegraphics[width=0.5\textwidth]{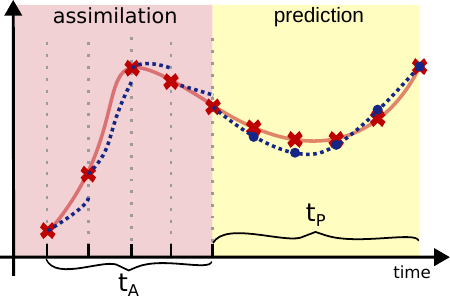} 
    \caption{Schematic representation of the prediction process. The system is trained to predict the trajectory over a time window $t_P$ based on a set of measurements. Assimilation occurs during a time window $t_A$, in which the system state is reinitialized with the available measurements (red markers in the figure). The figure also shows the network's predictions at the measurement times (blue dots).}
    \label{fig:traj_scheme}
\end{figure}

\section{Hybrid modeling and time series prediction}

To address the challenge of predicting trajectories in chaotic systems, we propose a hybrid model that combines differential equations with recurrent neural network. 

Given a set of measurements $\{ \vect{y}_i \}_{i=0}^{T_M}$ sampled from a trajectory $\vect{x}^{R}(t)$ - solution of a set of differential equations \eqref{eq:DS} with parameters $\vect{q}=\vect{q}_{R}$ - we assume that the observation provides a complete knowledge of the state $\vect{x}$ at the measurement time. Additionally, we consider measurements to be equally spaced in time with a time step $t_{\tau}= \tau \text{d}t $, where $\tau$ integer and $\text{d}t$ represents the integration time step. Then: 
\[
 \vect{y}_i \equiv \vect{x}^{R}(t=it_{\tau}) + \vect{\sigma}_{O},\]
where $i = 0, 1, \dots, T_M$ and $\vect{\sigma}_{O}$ quantifies the uncertainty due to the measurement process.

From the observations, we can estimate the unknown parameters $\vect{q}_{R}$ \cite{bagnoli2023synchronization}. However, due to the chaotic nature of the system, these estimates - even with complete knowledge of the state of the system $\vect{y}_i$ - are not sufficient for accurate future predictions. Furthermore, we cannot assume that we know the parameters of the system with arbitrary precision. In general, we can write that $\vect{q}_{M} = \vect{q}_{R} + \vect{\eta}$, where $\vect{\eta}$ denotes the error associated with the estimated parameters, which will depend on the method chosen to find the parameters and on the precision of the set of measurements $\{\vect{y}_i \}_{i=0}^{T_M} $.

The goal is that of predicting next states of the system using this set of measurements. To achieve this, we propose a hybrid method that combines general dynamics information with available local measures (Fig.~\ref{fig:traj_scheme}).\\

\subsection{Predictive framework}

\begin{figure}
    \centering
    \begin{tabular}{c c}
        (a) & \includegraphics[width=0.45\textwidth]{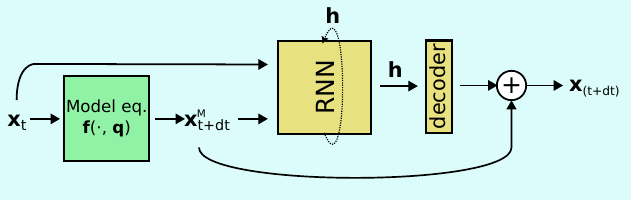} \\
        (b) &     \includegraphics[width=0.7\textwidth]{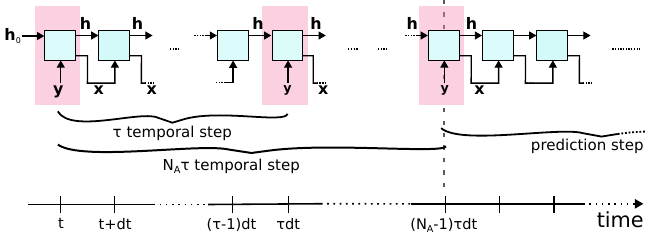}
    \end{tabular} 
    \caption{(a) Architecture used for a single time step integration. First we use the model equation to estimate $\vect{x}^M(t+\text{d}t)$, the prediction of the new state starting from an initial condition $\vect{x}(t)$. This two  are used as input of the neural network to  estimate the new state $\vect{x}(t+\text{d}t)$, using a residual approach.\\
    (b) Loop to generate a trajectory starting from $N_A$ measurement. Given an initial condition $\vect{x}_0 = \vect{y}_{t_0}$ and the initial hidden state $\vect{h}_0 = \vect{0}$, the system is evolved to predict the new states $\vect{x}_{\text{d}t}, \vect{h}_{\text{d}t}$. This values are used as the new input state for the next iteration, except when a measurement is available in the assimilation window (corresponding to time marked in pink in figure), where the measurement $\vect{y}_t$ is used as $\vect{x}_t$ input. The cyan boxes represent the architecture described in (a) in its unrolled representation.}
    \label{fig:architecture}
\end{figure}

The hybrid system integrates predictions from a physical model, denoted as \( \vect{f} \), with corrections generated by a neural network. Given a state \( \mathbf{x}(t) \) at time $t$, to estimate the new state at time \(\vect{x}(t + \text{d}t)\) the evolution of the system is computed according with the following sequence of operations:
\begin{itemize}   
    \item \textbf{Physical model prediction}: The differential equation \( \vect{f} \) is used to compute a preliminary estimate of the state at the next time step, \( \mathbf{x}^M(t + \text{d}t) \). The numerical integration scheme employed (e.g., Euler, Runge-Kutta, or Lsoda) determines the specific computation.
    
    \item \textbf{Neural network correction}: A neural network is used to estimate the correction term \( \boldsymbol{\xi}(t) \) using the initial condition and the estimated prediction as input.

    \item \textbf{State update}: The predicted state in the next time step is obtained by combining the estimate from the physical model with the correction from neural network:  
    \[
        \mathbf{x}(t + \text{d}t) = \mathbf{x}^M(t + \text{d}t) + \boldsymbol{\xi}(t).
    \]
    This estimated state is used as input for the subsequent time step t + \text{d}t.
\end{itemize}
Starting from the initial condition $\vect{x}(t=t_0)$ and the state  $\vect{x}^M(t=t_0 + \text{d}t)$ predicted by the model $\vect{f}(\cdot,\vect{q}_{M})$, the network is trained to correct the model's output, enabling long-time predictability.

\subsubsection*{Architecture}
The situation described in the previous section 
suggests using a RNN, a neural network architecture with temporal memory, which provides flexibility in capturing dynamics with temporal dependencies.

The proposed neural network architecture consists of two main components (Fig.~\ref{fig:architecture}(a)):
\begin{itemize} 
    \item \textbf{Memory block}. Encodes and corrects the trajectory in a high-dimensional latent space of dimension $H_L$.
    \item \textbf{Decoder}. Comprises two linear layers of dimension $H_L$, followed by a ReLU function to introduce a non-linearity, and a output linear layer to remap the network state back into the system's space.
\end{itemize}

For the memory block we use the GRU architecture. 
It evolves an internal state $\vect{h}(t)$ combining present and past information through two gates, which determine the relevance of past trajectory information to preserve for the next state.
Typically, and also in this work, the state $\vect{h}_0 = \vect{h}(0)$ is initialized to zero, but the impact of further initializations has been analyzed in the literature \cite{analisysH0state}.

\subsubsection*{Training and interpolation}
To contextualize the problem within a short-term forecasting framework, the measurements are organized in windows of fixed length \( N_W = N_A + N_P\), where \( N_A \) and \(N_P\) represent the number of observations used for the assimilation and the prediction steps.

A single measurement window covers a time span of length~\( \Delta t_W = (\tau N_W+1)\text{d}t \). This time span is subdivided into two phases: assimilation and prediction (Fig~\ref{fig:traj_scheme}). During the assimilation window, the data are used to correct the state estimated by the neural network module, while in the prediction phase, the network evolves freely without the use of measurements.

The data set is therefore made up of \textit{measurement windows} 
\[ \vect{Y}_i = \{\vect{y}_i, \vect{y}_{i+1}, \dots, \vect{y}_{i+N_W}\} = \{\vect{y}_j\}_{j=i}^{i+N_W} \]
generated from the observed time series $\{\vect{y}_i\}_{i=0}^{T_M}$.

In the following results, the training data consists of time series generated from the evolution of the system \( \vect{f} \) with parameters \( \vect{q}_{R} \),  using the Runge-Kutta45 method.  It is assumed that the full system state is observed every \( \tau \) time steps \( \text{d}t \). 
To simulate realistic conditions, white random noise in the range \( \vect{\sigma}_{O} \) or Gaussian noise whit standard deviation \( \vect{\sigma}_{O} \) was added to the data after trajectory generation. In both cases the hybrid system shows similar results.

The dataset is split into three subsets (training = 64\%, validation = 16\%, and test = 20\%) without temporal shuffling, ensuring that the test set consists of the last time steps. This approach provides a more reliable estimation of future prediction errors and reduces potential overfitting.

During training, the network must predict a time window \( t_{P} = N_P t_{\tau} \) starting from  $N_A$ measurement $\{\vect{y}_j \}_{j=i}^{i+N_A}$ in the assimilation window of length \( t_A = N_A t_{\tau} \).

Given a measurement set \( \vect{Y}_i \), the system is initialized with the fist measurement \(\vect{x}_0^{(i)} = \vect{x}(t=0)^{(i)}= \vect{y}_{i}\), where the index $i$ in $\vect{x}^{(i)}$ corresponds to the trajectory generated from the i-th measurement window, and performs a predictive step with the model system \( \vect{x}^M_{\text{d}t} \). The input and model estimate are fed into the network, and the prediction at the next time ($\vect{x}_{\text{d}t}^{(i)}$) is a combination of the network output and the first model output \( \vect{x}^M_{\text{d}t} \). 

To enforce adherence to the true trajectory during training, the system uses the first \( N_A \) measured states in each window directly as inputs at the corresponding time steps. This \textit{assimilation phase} ensures that the hybrid model aligns closely with the observed dynamics over this interval. In the \textit{prediction phase}, which spans the subsequent \( N_P \) observations, the predicted states generated by the hybrid system are fed back as inputs for future time steps, emulating autonomous evolution. During training, the weights of the neural network are optimized by minimizing the error between the predicted states and the true measurements in the prediction window. 

The cost function used for the training is the $L_2$ norm over the difference between the observations and the hybrid system's outputs at the measurement time in the prediction window:
\begin{equation}
    \mathcal{L} = \sum_{i} || \{\vect{y}_j \}_{j=i+N_A}^{i+N_W} - \{ \vect{x}_{jt_{\tau}}^{(i)}\}_{j=N_A}^{N_W}||_2,
    \label{eq:loss}
\end{equation}
where $\{\vect{y}_j \}_{j=i+N_A}^{i+N_W} $ denote the observation in the prediction time $t_P$ for the i-th measurement window,  $\{\vect{x}_{j t_{\tau}}^{(i)}\}$ are the corresponding prediction of the hybrid system at the measurement time $jt_{\tau}$, for $j=N_A, N_A +1, \dots, N_W$, and 
the index $i$ runs over the elements $\{\vect{Y}\}_i$ considered in the training.
The general algorithm is schematized in Fig.~\ref{fig:architecture}(b).

We would like to point out that the system is explicitly designed to handle measurements available only at discrete intervals of \( t_{\tau} = \tau \text{d}t \). By combining the neural correction with the physical model integrated at temporal step $\text{d}t<t_{\tau}$,  the hybrid system can also interpolate the state evolution at intermediate time steps where direct measurements are unavailable.

\subsubsection*{Handling partial measurements}
\label{sss:PartialMeasure}
In practical scenarios, not all system variables can be measured. Often, only a subset of variables is accessible or convenient to measure for a long time. To address this challenge, our model must be capable of handling partial measurements.

To predict new states using the model Eq.~\eqref{eq:DS} a complete knowledge of the initial conditions are typically required unless there is a direct and known dependency between the measured and unmeasured variables.
However, if our hybrid model has correctly learned the dynamics, we can use a genetic algorithm approach to estimate the full system state from a set of partial measurements.

For example, suppose only one variable, such as the \( x \) direction, can be measured. Starting from the initial condition \( x(t) \), we generate an ensemble of $M$ replica of our system, initialized with the observation in the measurement direction, while random values are assigned to the unknown directions based on the probability distributions of the corresponding variables in the model $\vect{f}$.

This ensemble of initial conditions is then evolved over time. During the assimilation phase, whenever a measurement is available, a \textit{pruning-enriching} procedure is performed. In order to favor predictions closer to the measured trajectory, states with higher distances between the measurement and the simulated value are replaced by noisy copies of states with the lowest distance. 
In that way the ensemble of initial conditions is forced to follow the most probable trajectory given the partial measurement and generates a beam of trajectories that follow the trend of the observed trajectory. Instead, during the forecasting phase the ensemble is left free to evolve.

Practically, we initialize $M$ ensemble elements and evolve them for a time interval $t_{\tau}$, when the next measurement is available.
We then sort the ensemble elements in ascending order based on the distance between the measured direction and the corresponding ensemble predictions, replacing the farthest ensemble members with copies of the first half.
This procedure is iterated throughout the assimilation window, after which the elements of the ensemble are free to evolve independently.

To increase the variability of the response and allow exploration of space near the ensemble elements, the copies are added with noise equally distributed in [$- \nu$, $ \nu$], otherwise the whole ensemble degenerates to the same trajectory after few pruning-enriching steps.
The amplitude of $\nu$ will determine the weight of this exploration, so it is very important to select a sufficient large values of  $ \nu$, but in relation to the typical size of the attractor, that can be estimated by the data and the dynamical system $\vect{f}$. 

The predicted trajectory is the average of the first $M_e$ elements of the ensemble. This allows us to define an uncertainty on the prediction based on the standard deviation on the first $M_e$ elements of the ensemble.

\section{Experimental evaluation}
\subsubsection*{Experimental configuration}
The neural network in the hybrid model does not require much computational effort, at least in the low-dimensional cases considered in this study. The calculation time required may depend more on the $\vect{f}$ model considered and the integration scheme used. In the cases we will present in the next section, the initial prediction $\vect{x}_{M}$ was obtained by numerically integrate the dynamical system $\vect{f}$ using the first order Euler scheme. However, in the presence of unstable dynamics that require more precise algorithms, the choice of calculation speed and performance trade-off will be crucial and will therefore require further in-depth studies.

The network training was performed in a Torch environment \cite{paszke2019pytorch} with a 12 GB NVIDIA RTX 5000 GPU (but the maximum memory requirement for the systems analyzed in the described training conditions is on the order of 2 GB). The optimization was performed using the Adam algorithm with learning rate $l_r = 10^{-3}$ and a $L_2$ regularization of amplitude $10^{-4}$. The network includes about 80K parameters and the system was trained for 300 epochs. For the chosen hyperparameters and the dynamical models considered in the study, typical training times are in the order of two hours.

\subsubsection*{Quality indicator of the results}
To evaluate the performance of each model we use the root mean square error, $E(t)$, as an indicator of the distance in the state space, averaged over all the measurement windows of the test set. 
This indicator allows us to quantify the amplification of the prediction error in the prediction window. From this it is possible to obtain two scalar indicators: the $\langle E \rangle$, that is the $E(t)$ averaged over the window, and its maximum value in the window, $E_{MAX}$.

The mathematical definitions are as follows.

For the $E(t)$ we have:
\begin{equation}
    E(t) = \sqrt{\frac{1}{N} \sum_i ||\vect{y}_i(t) - \hat{\vect{y}}_i(t)||_2},
    \label{eq:E}
\end{equation}

where the index $i$ runs over the different window prediction $\{\vect{Y}_i\}$ in the test set, $\vect{y}_i$ and $\hat{\vect{y}}_i$ are, respectively, the true and the predicted value of the state in the prediction window, and the operator $||\cdot||_2$  indicates the $L_2$ norm in the state space.

The $\langle E \rangle$ and the maximum value $E^{\text{MAX}}$ are, respectively, the average over the times $t_i$ in the the window prediction of the $E(t)$:
\begin{equation}
    \langle E \rangle = \frac{1}{N_i} \sum_i E(t_i),
    \label{eq:Emean}
\end{equation}
and the max value of $E(t)$ over the same times:
\begin{equation}
    E^{\text{MAX}} =  \max_{i}   E(t_i),
    \label{eq:Emax}
\end{equation}
where the index $i$ is a time index that runs on the time values $t_i$ used to compute the $E(t)$.

In the definitions we have left ambiguity of the times $t_i$ on which to perform the average and max operations since the predictions corresponding to the measurement times are only a fraction of the predictions made by the hybrid system in the entire prediction window, depending on the chosen $t_i$ values we can define two average indicators.

The value of $E(t)$ averaging (its maximum) on only the observation times: $\langle E \rangle_O$ ($E^{\text{MAX}}_O$) quantifies the prediction error of the hybrid model in the measurement times, and this is the only average indicator that can be used in real situations.

To quantify the learning of the hybrid system at the intermediate measurement times, and thus understand the capabilities of the hybrid model to correct interpolations between measurements, we also define the $\langle E \rangle_P$ ($E^{\text{MAX}}_P$), which is the $E(t)$ averaged (maximum) over all prediction times.

\section{Results}
We tested the procedure on low-dimensional dynamical systems, in particular the Lorenz '63 and Lorenz '83 models, also known as the low-dimensional Hadley model and the R\"ossler system~\cite{rossler1976equation}. In the following, we will report the graphical results for the Hadley circulation model and the Lorenz '63 system.

In both cases the system was initialized in a random state and left to evolve for a time \( T=500 \). After a transient time \(T_{\text{trans}} = 40 \), to ensure that the system is on the attractor, we selected measurements every \( \tau \) time steps with \( \text{d}t = 0.005 \). In the results will be show below, to emulate measurement situations, white random noise with an amplitude of the order of $5 \%$ of the typical range of the system variables was added to the generated measurements.

\subsubsection*{Hadley circulation}
The Hadley circulation model, in this low-dimensional form, was introduced by Lorenz in 1983 \cite{lorenz1984irregularity} following studies on global circulation. 
This model explores the dynamics of the global westerly wind current and its interactions with large-scale atmospheric eddies. 
Looking for a simple but effective description of the problem, reducing it to the main degrees of freedom, Lorenz defines the following set of three-dimensional differential equations:  
\begin{figure}
    \centering
    \includegraphics[height=0.15\textheight]{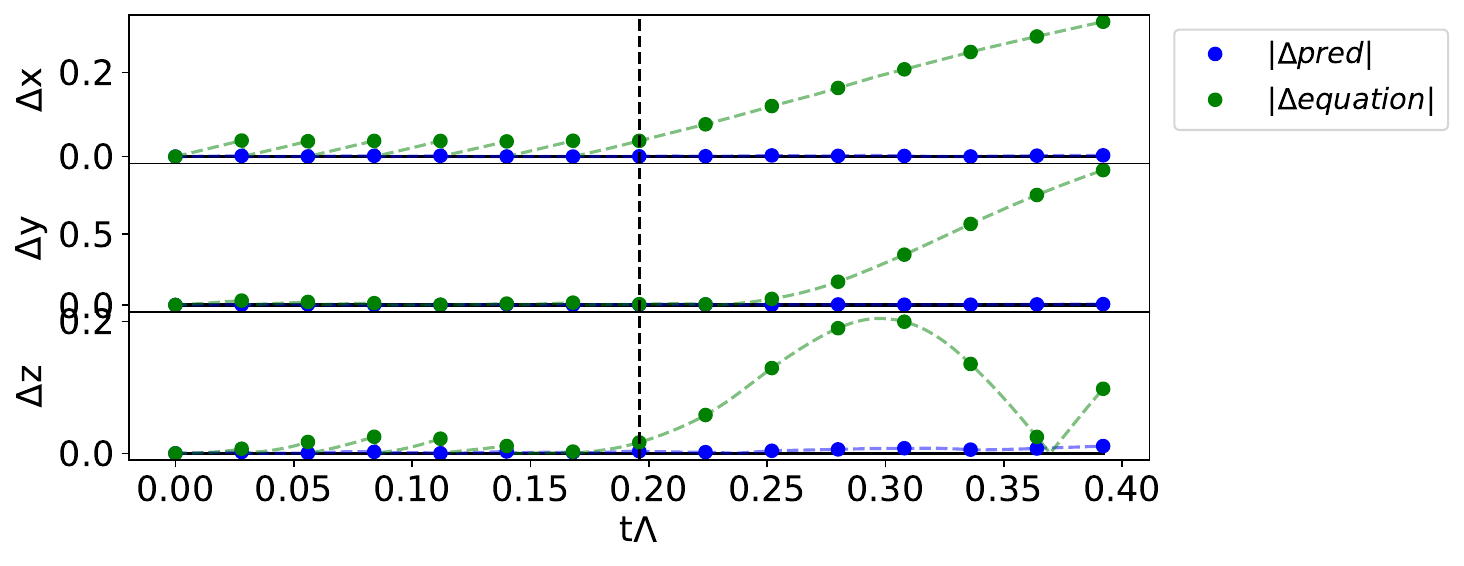} \\
    \caption{Example of network error prediction with respect to the true trajectory starting from a set of initial conditions $ \{ \vect{y}_{t_i} \}_{i=0}^{N_A} $. We show the error between the predicted trajectories and the observations ($\Delta \text{pred}$ in blue for the prediction with the hybrid system, $\Delta \text{equation}$ in green for the trajectory generated with only the dynamical system described by the function $\vect{f}(\vect{x}, \vect{q}_M)$).}
    \label{fig:PredHadley}
\end{figure}
\begin{equation}
    \label{eq:hadley}
    \begin{split}
        \dot{x} &= - a x + a F - y^2  - z^2,\\
        \dot{y} &= -y + G + xy - bxz,\\
        \dot{z} &= -z + xz + bxy. 
    \end{split}
\end{equation}

Where $\vect{q} = (a, b, F, G)$ are the parameters of the problem.

We initialize the system with real parameters: $\vect{q}_{R} = ( 0.25, 4, 8, 1 )$, values for which the system exhibits chaotic behavior, and a noise amplitude $ \eta=0.2$  for the parameters.
As for the sampling of measurements, we assume that they are accessible every $ \tau=20$ time step $\text{d}t$, with a time between two measurements $t_{\tau} = 0.1$ in the time units of the system. The results shown were obtained using a measurement window with $N_W = 15$ measurements, of which $N_A=7$ where used the assimilation phase, while the forecast window consist of $N_P=8$ observations.

In Figure~\ref{fig:PredHadley} we show the prediction error of a single measurement window. For comparison, we also show the results of a prediction starting from the last assimilated measurement obtained from the dynamic evolution of the equation~\eqref{eq:DS}  with the estimated parameters $\vect{q} = \vect{q}_{M}$. 
As can be seen, in this forecast window the trajectory predicted by the dynamic equation $\vect{f}(\cdot, \vect{q}_{M})$ shows divergence due to the imperfect knowledge of the parameters, which amplifies the intrinsic divergence given by the chaotic nature of the system.
The modification made by the network to the model estimate at each time step allows it to constantly correct the forecast and synchronize the output with the true trajectory.
\begin{figure}
    \centering
    \includegraphics[width=0.8\textwidth]{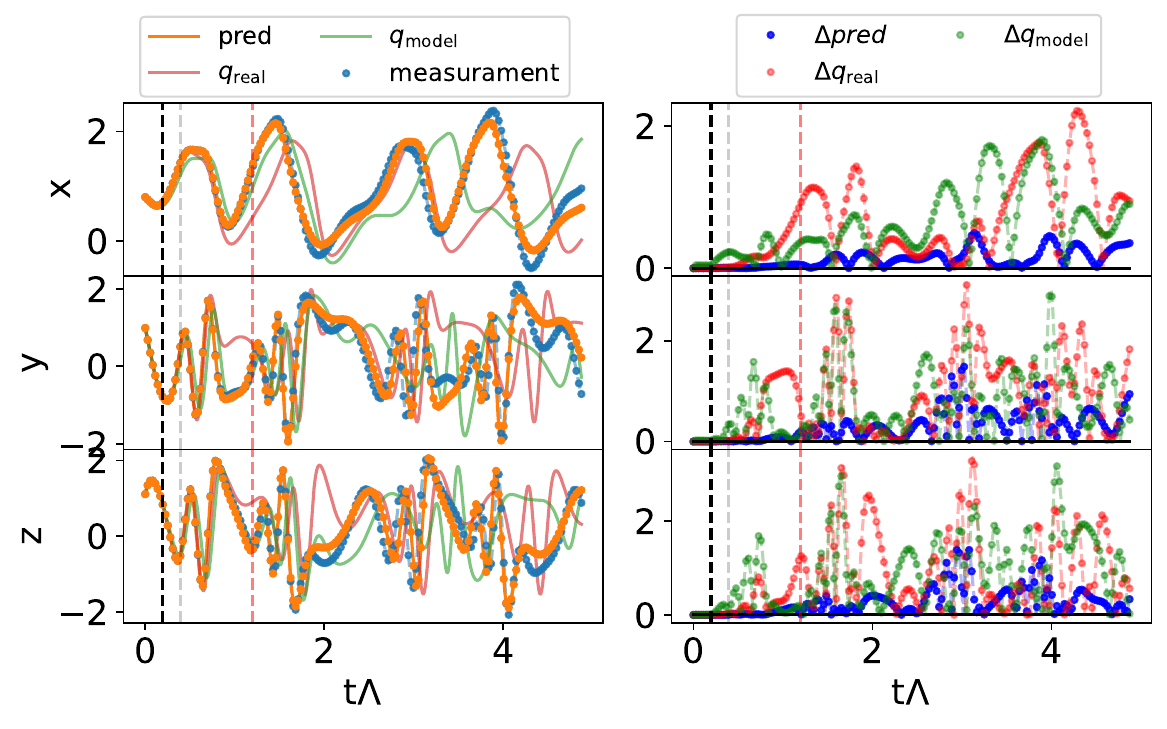}
    \caption{Prediction from an initial condition (point above the black vertical line). On the left, we show the prediction of the hybrid system (pred in the legend), and those obtained from the dynamics equations for the different parameters. In blue the real measures. On the right instead,  we show the prediction error in the different cases. The time used as a prediction window in the training phase is marked with the gray vertical line. The red line indicates the Lyapunov time of the system.}
    \label{fig:CIHadley}
\end{figure}

Let us now analyze the prediction results from the initial conditions, specifically the first $N_A$ observations $\{\vect{y}_i\}_{i=0}^{T_A}$, for a prediction time longer than the one used in the training phase, $t_P$. 

In Fig.~\ref{fig:CIHadley}, in the left column, we show the prediction of the hybrid system when it is left free to evolve for a time $t_P'\gg t_P$, along with the absolute value of its error prediction (right column). For comparison, we also show the prediction from observations of the dynamical system $\vect{f}$ both in the case of real parameters $ \vect{q}_{R}$ and for the parameters used in the equation developed by the hybrid model, $ \vect{q}_{M}$. As we can see, the trajectory predicted by the hybrid model allows for the correct estimation of the long-term behavior of the chaotic trajectory, even if it does not follow the real state exactly.
\begin{figure}
    \centering
    \begin{tabular}{c c}
         \includegraphics[height=0.18\textheight]{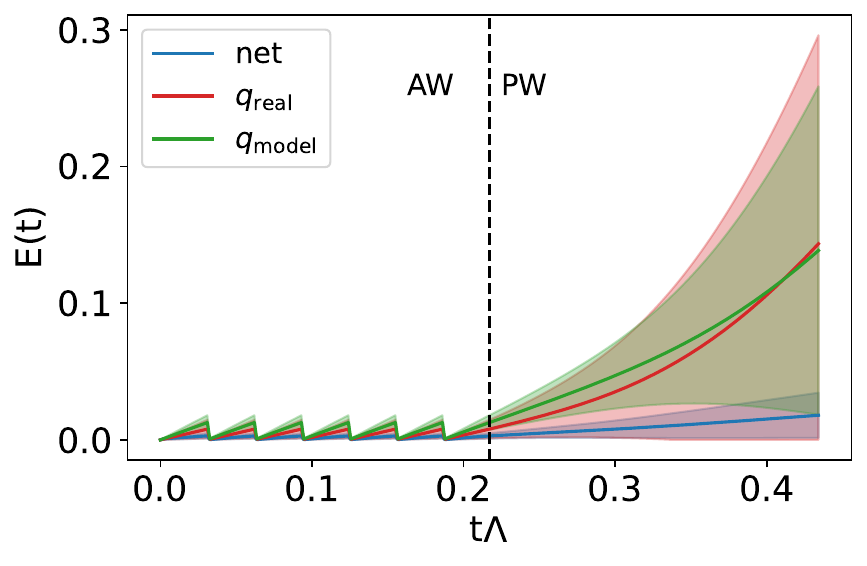} & \includegraphics[height=0.18\textheight]{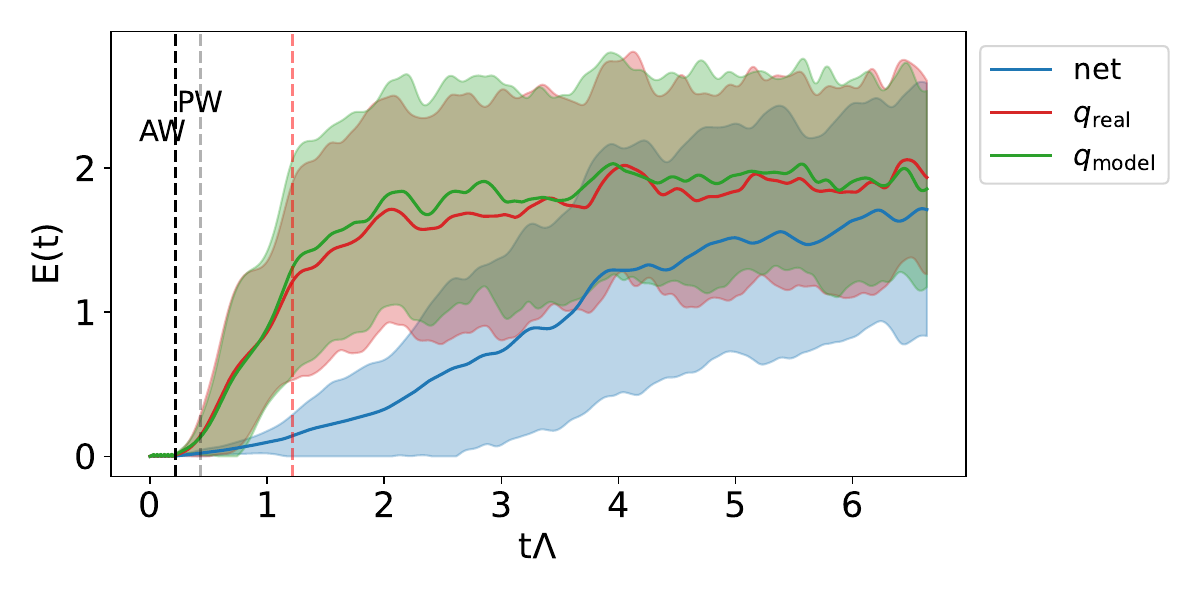} \\
         (a) & (b) \\
    \end{tabular}  
    \caption{$E(t)$ and its variance for a prediction time equal to that used in the training procedure (a) and for a longer prediction time (b). In panel (b), with the gray vertical line we indicate the end of the prediction window considered during the training, with red vertical line we show the Lyapunov time of the system (starting from $t_p$).}
    \label{fig:averageErrorWondowHadley}
\end{figure}

A better understanding of the effectiveness of this model and the contribution it can give to the prediction of chaotic systems can be gained by observing the average error in the prediction window in different cases; the hybrid model and the differential equation with different parameters ($\vect{q}_{R}, \vect{q}_{M}$).

In Figure \ref{fig:averageErrorWondowHadley}, we show the indicator $E(t)$ calculated according to Eq.~\eqref{eq:E} on $100$ randomly extracted initial condition in the test set in the following cases: (a) a precision window corresponding to that used for the training phase, and (b) a prediction time twenty times larger, along with its standard deviations.

As shown in Fig.~\ref{fig:averageErrorWondowHadley}(a), the trajectories generated by the dynamic model exhibit the expected exponential divergence, which is more rapid in the case of \(\vect{q}_{M}\) partly due to parameter mismatching. Conversely, the hybrid model predictions follow the true trajectory, diverging over time but at a significantly lower rate than that observed for the standalone models. This behavior persists even over long time horizons (Fig.\ref{fig:averageErrorWondowHadley}(b)), where the Lyapunov time — defined as the inverse of $\Lambda$, the maximum Lyapunov exponent of the system — is also indicated with a dashed red line. While the trajectories predicted by the models show a distance from the true trajectory comparable to the spatial scale of the attractor, the hybrid model achieves relative errors on the order of \(20\%\).
\begin{table}
\centering
\begin{tabular}{c  c c | c c | c c}
    \toprule
    \multirow{2}{*}{$E$} &
    
    \multicolumn{2}{c|}{Hadley}& \multicolumn{2}{c|}{Lorenz63} & \multicolumn{2}{c}{R\"ossler} \\ [+0.0em]

    &  $\scriptstyle \langle \cdot \rangle$ & $\scriptstyle {MAX}$ & $\scriptstyle \langle \cdot\rangle$ & $\scriptstyle MAX$ & $\scriptstyle \langle \cdot \rangle$ & $\scriptstyle MAX$\\ 
    \hhline{=======} \\[-1.0em]
    
    \multicolumn{7}{c}{observation (O)}\\
    \hline \\[-1.0em]
    
    hybrid model                        & 0.0057 & 0.0104 & 0.5492 & 0.7778 & 0.0092 & 0.0140\\
    $\vect{f}(\cdot, \vect{q}_{R})$  & 0.0347 & 0.0828 & 3.0782 & 6.5586 & 0.0309 & 0.0536\\
    $\vect{f}(\cdot, \vect{q}_{M})$ & 0.1164 & 0.2532 & 3.1209 & 6.5061 & 0.1624 & 0.3161\\
    \hhline{=======} \\[-1.0em]
    
    \multicolumn{7}{c}{all prediction window (P)}\\
    \hline \\[-1.0em]
    
    hybrid model                        & 0.0057 & 0.0104 & 0.5563 & 0.7778 & 0.0092 & 0.0140\\
    $\vect{f}(\cdot, \vect{q}_{R})$  & 0.0333 & 0.0828 & 3.0285 & 6.5586 & 0.0309 & 0.0536\\
    $\vect{f}(\cdot, \vect{q}_{M})$ & 0.1136 & 0.2532 & 3.1476 & 6.6616 & 0.1617 & 0.3161\\
    \bottomrule
\end{tabular}  
\caption{Average and maximum prediction error, calculated over the entire test set. In the first part of the table we show the indices averaged over the times corresponding to the measurement times, in the second part instead those averaged over the entire prediction window $t_P$.}
\label{table:1}
\end{table}

In Table~\ref{table:1}, we show the scalar indicators defined in Eq.~\eqref{eq:Emean} and Eq.~\eqref{eq:Emax}, calculated with respect to the measurement times (top of the table) and with respect to the entire forecast window (bottom of the table).

As can be observed, the correction at each time step d$t$ made by the network allows it to correct and force the predictions of the hybrid system to follow that observed, resulting in an up to 100 times improvement in the mean accuracy of short-term prediction compared to the solutions provided by the evolution of the dynamics $ \vect{f}$.

Furthermore, we observe that, on average, there are no significant differences between the predictions corresponding to the measurement times, on which the optimization process was based, and those averaged over the entire prediction window. This indicates that the system has correctly learned the corrections to be made to the local dynamics of the system, without however modifying it profoundly. The dynamic model defined within the hybrid system drives the global dynamics, the network corrects the divergences, attenuating them and forcing the prediction to follow the real trajectory.
\begin{figure}
     \centering
    \begin{tabular}{c c c}
         (a) & \includegraphics[height=0.1\textheight]{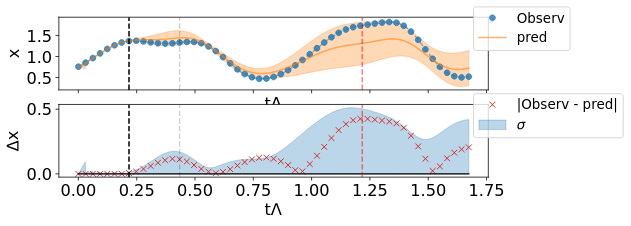}  & \includegraphics[height=0.1\textheight]{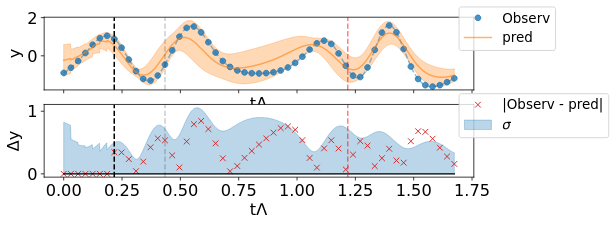} \\
         (b) & \includegraphics[height=0.1\textheight]{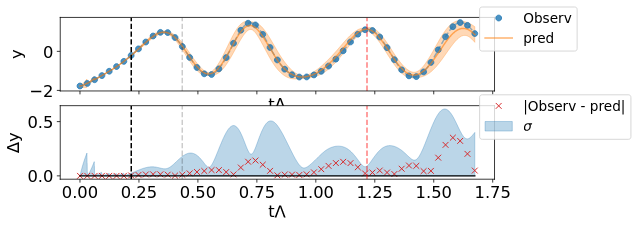} & \includegraphics[height=0.1\textheight]{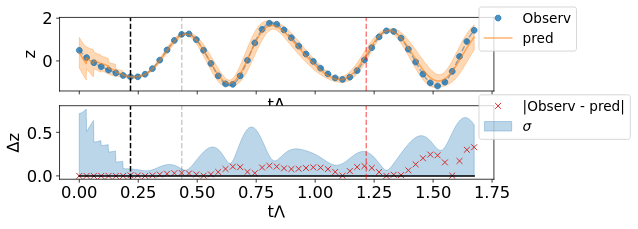} 
    \end{tabular}  
    \caption{Prediction from partial measurements using an ensemble of $M=256$ elements. (a) Measurements only in the $x$-direction and (b) only in $y$-direction. For each plot we show, in the top panel, the estimated trajectory as mean of the first $M_e=64$ elements of the ensemble with the estimated error in orange and in blue the true measurement. In  the bottom panel, we show the prediction errors (x markers) and the standard deviation $\sigma$ of the ensemble prediction. In the plots we also show the end of the assimilation window (black vertical line), the end of the prediction window used in the training phase (gray vertical line) and the Lyapunov time (red vertical line).}
    \label{fig:PartialMisHadley}
\end{figure}

Let us now analyze the predictions in the case of partial measurements. In particular, in Figure~\ref{fig:PartialMisHadley} we show the prediction starting from the initial condition $ \{ \vect{y}_{t_i} \}_{i=0}^{N_A} $, assuming that the system is measured only in one direction. In Figure~\ref{fig:PartialMisHadley}(a) we suppose that only the  $y$ direction is measured, in Figure~\ref{fig:PartialMisHadley}(b) the $x$ direction is considered as direction of measure. 
The generation was obtained from an ensemble of $M = 256$ elements, using amplitude noise at the pruning step $\nu=0.2$.
The predicted trajectory and its standard deviation are obtained from the first $M_e=64$ elements of the ensemble.
In Figure~\ref{fig:PartialMisHadley} we show two different predictive scenarios. In particular in Figure~\ref{fig:PartialMisHadley}(a), a low uncertainty is observed in the standard deviation of the ensemble prediction in the measurement direction, which is reflected in a low standard deviation also in the predicted directions. Inversely from what is observed in Figure~\ref{fig:PartialMisHadley}(b), where a large uncertainty in the measurement direction, is reflected in a large uncertainty in the prediction of the hidden directions.

However, these results are also affected by the choice of the elements of the ensemble $M$ and by the number $M_E$ of elements used to obtain the average prediction. Further studies will be needed to understand and quantify the error statistics in this predictive scenario.


\subsubsection*{Lorenz 63}
The Lorenz 63 model, introduced by Edward Lorenz in 1963 \cite{DeterministicNonperiodicFlow}, is a simplified mathematical model designed to study convection in an atmospheric layer in the presence of a temperature gradient.

The system is described by three coupled, nonlinear ordinary differential equations:
\begin{equation}
    \label{eq:lorenz}
    \begin{split}
        \dot{x} &= \sigma (y-x),\\
        \dot{y} &= -xz + \rho x -y,\\
        \dot{z} &= xy -\beta z. 
    \end{split}
\end{equation}




We considered measurement windows composed of $N_A=7$ and $N_P=8$ and $\tau = 30$, so each measurement is available every $t_{\tau} = 0.15$ in the time units of the system.

\begin{figure}
    \centering
    \begin{tabular}{c c}
         \includegraphics[height=0.18\textheight]{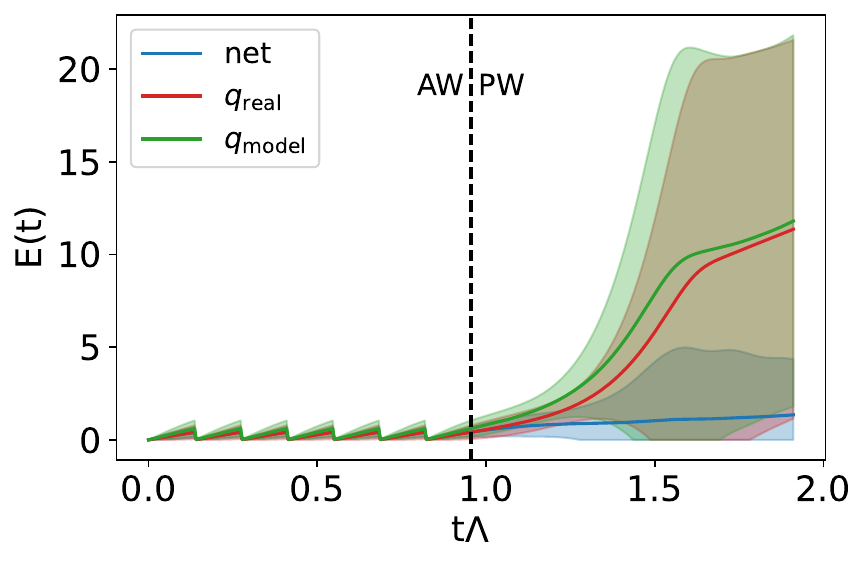} & \includegraphics[height=0.18\textheight]{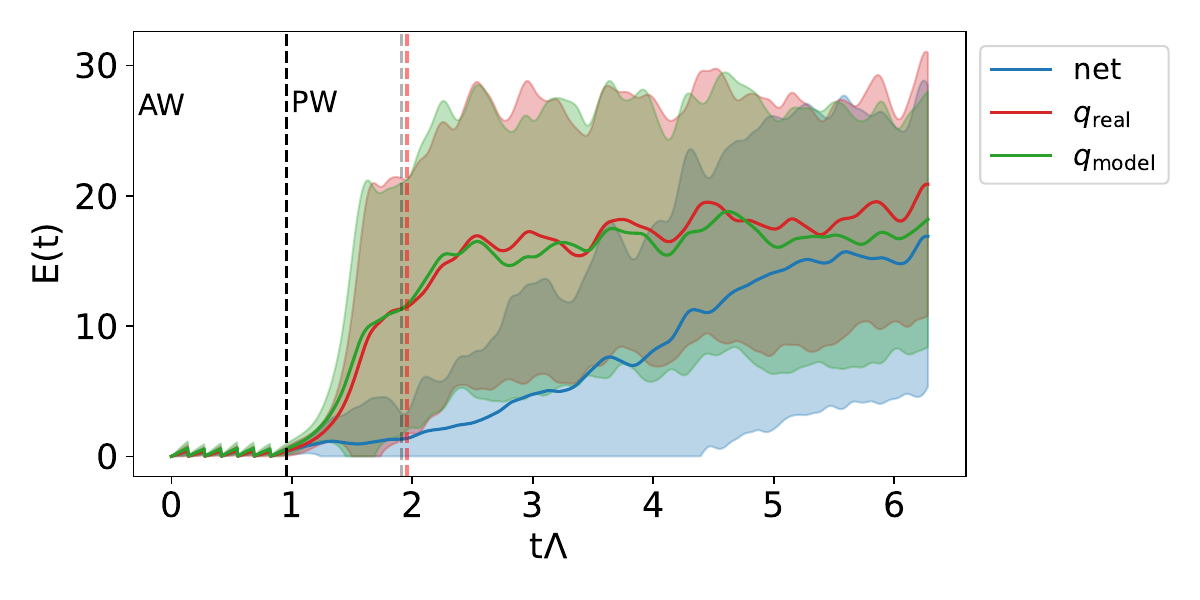} \\
         (a) & (b) \\
    \end{tabular}  
    \caption{$E(t)$ and its variance for a prediction time equal to that used in the training procedure (a) and for a longer prediction time (b). With the red vertical line we show the Lyapunov time of the system.}
    \label{fig:averageErrorWondowLorenz}
\end{figure}

Below we show the results of the average error predictions and of forecasting starting from a set of measurements.
As for the previous experiment, in Figure \ref{fig:averageErrorWondowLorenz}, we show the indicator $E(t)$ calculated according to Eq.~\eqref{eq:E} on $200$ randomly extracted initial condition in the test set for short (a) an long time prediction (b), along with its standard deviations.
Again, the correction provided by the hybrid system allow to generate trajectory that follow the true one for a longer time, compared to the Lyapunov time of the system.

\begin{figure}
    \centering
    \begin{tabular}{c c}
        (a) & \includegraphics[height=0.13\textheight]{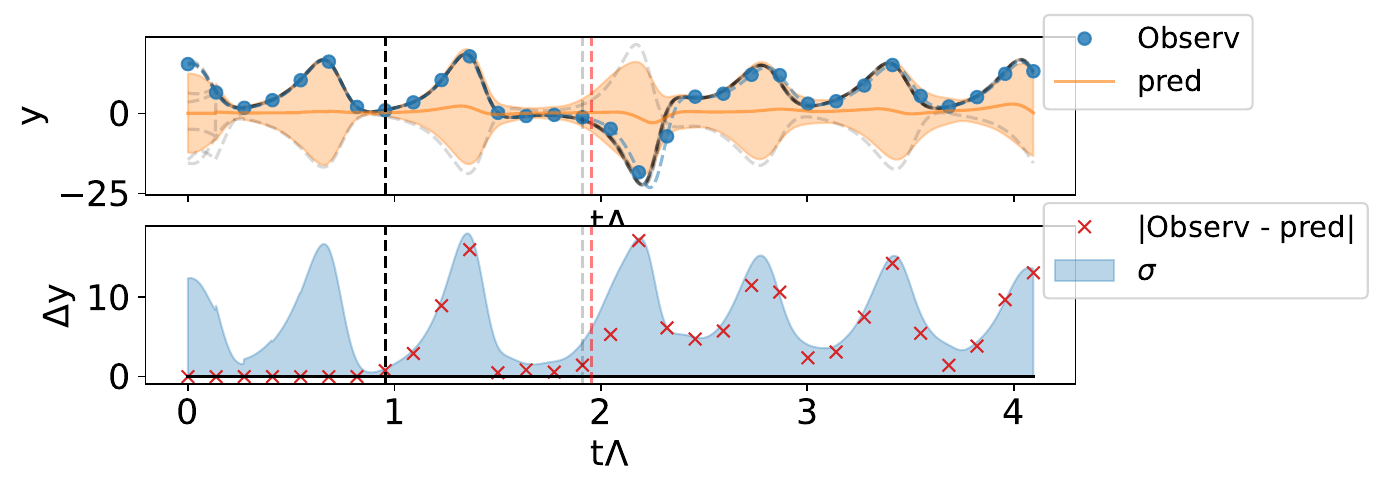} \\   
        (b) & \includegraphics[height=0.13\textheight]{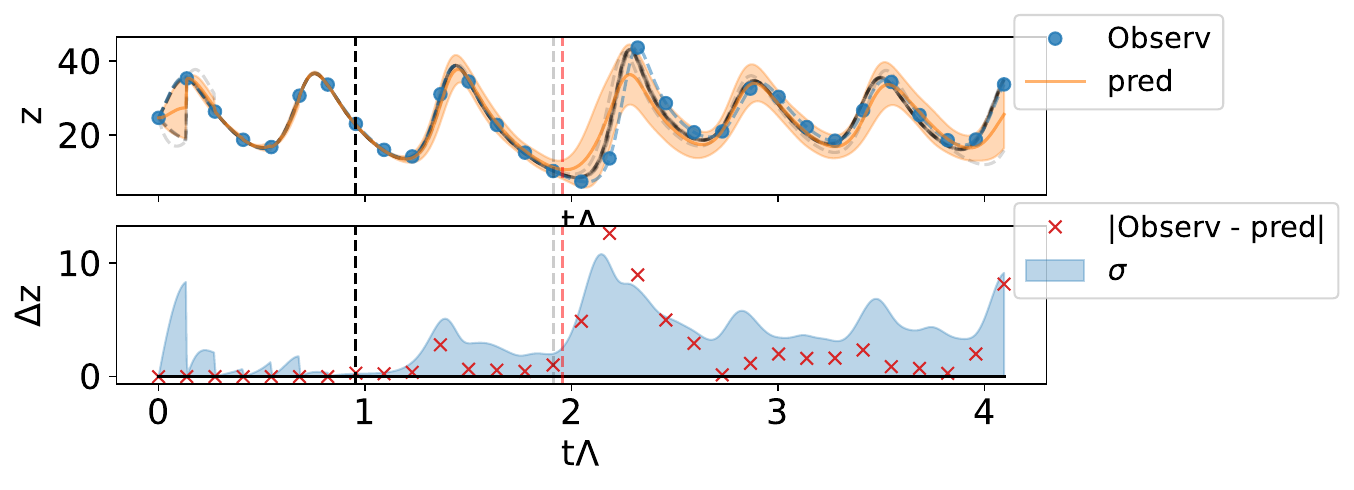}
    \end{tabular}  
    \caption{Prediction from partial measurements using an ensemble of $M=1024$ elements - Measurements only in the $z$-direction. (a) $y$-direction, (b) $z$-direction. See the caption of Fig.~\ref{fig:PartialMisHadley} for a full description of the meaning of the elements shown in the figure. In gray line we also show the trajectories of the first 10 elements of the ensemble (see text).
}
    \label{fig:PartialMisLorenz}
\end{figure}

In Figure \ref{fig:PartialMisLorenz}, we show the case of ensemble prediction, assuming that only one variable is measured as initial conditions. In particular, we show the behavior when the variable $z$ is assimilated. It is well known in the literature that this variable is not a good synchronization direction \cite{Pecora2015}, and here, for assimilation: the information on the direction $z$ does not uniquely identify the position of the system on the attractor, due to the symmetry of the system with respect to the transformation $(x, y, z) \rightarrow (-x, -y, z)$.
As we can see in Figure~\ref{fig:PartialMisLorenz}(a), the prediction of ensemble in the unobserved direction $y$ is unable to discriminate which of the trajectories symmetric with respect to the direction $y$ is the real one, showing an average behavior that is not typical of the dynamics of the system considered, even though the direction $z$ is determined within the errors (Fig.\ref{fig:PartialMisLorenz}(b)). To support this, we show the behavior of the first 10 trajectories of the hexamble (gray trajectories in the figure). As can be seen, while for the $z$ direction these are close to each other, and begin to diverge for prediction times greater than the system's launch time, the $y$ direction is characterized by symmetric $y$ and $-y$ trajectories.

This aspect shows how the architecture of the network does not break any symmetries of the system. Once again, we see that the main dynamics in this hybrid system is governed by the dynamic equation of the $\vect{f}_M$ model. The network learns the local corrections to be made over short times, but these are necessary to guarantee the convergence of the simulated trajectory on the observed data.


Assuming that we measure the $x$ or $y$ direction, the results obtained are analogous to those obtained using the Hadley model shown in the previous section.

\section{Conclusions}
In this work, we have presented a method to combine physical information with neural network techniques to predict future observations of chaotic systems and interpolate the states of the system between measurements.

After defining the hybrid model and presenting how to optimize it, we have shown the learning results for low-dimensional chaotic systems, demonstrating how the hybrid model is able to predict future states for both short times ($t_P$) and long times ($\gg t_P$), not only at the measurement times on which the system has been trained, but also for time values between two successive measurements.
Our system is therefore able to correctly interpolate even unknown states of the system.

Additionally, this hybrid approach shows promise in predicting the dynamics of systems with partial measurements. The hybrid model successfully synchronizes with the real system and corrects the simulated model, leading to improved predictions. However, as expected from theory, the trajectories begin to diverge exponentially according to the Lyapunov time, particularly for longer prediction intervals.

This work is a starting point for a more in-depth study aimed at analyze the use of artificial intelligence coupled with dynamical equations for the study of physical systems. Several considerations and future studies will need to be addressed. First, it will be important to analyze the impact of the measurement time $t_{\tau}$ chosen, to understand the influence of this parameter to the quality of the prediction of the hybrid system.

Furthermore, this approach will need to be extended to high-dimensional models to evaluate the impact of dimensionality on the quality of the results, as well as to spatially extended models, possibly by appropriately modifying the network structure to incorporate physical information in the architecture. For example, replacing the RNN with architectures that also include spatial correlations (e.g. ConvGRU or similar) could be a promising direction.

Finally, to make this system competitive and applicable to real-world scenarios, it is important to analyze the feasibility of this method in the presence of partial measurements, for example, by using pretrained hybrid models on synthetic data, or by modifying the form of the loss function (Eq.~\eqref{eq:loss}) to focus the optimization on the available measurements.

\printbibliography
\end{document}